\documentclass[11pt]{article}

\usepackage{amsmath}
\begin{document}

\title{\bf Symmetry Classification  of   quasi-linear PDE's Containing
Arbitrary Functions \\ ~\\}

\author{GIAMPAOLO CICOGNA \\ ~ \\
{\it Dipartimento di Fisica ``E.Fermi'' dell'Universit\`a di Pisa} \\
{\it and 
  Istituto Nazionale di Fisica Nucleare, Sez. di Pisa} \\ 
 {\it Largo B. 
Pontecorvo 3, Ed. B-C, I-56127, Pisa, Italy} \\
(fax: +39-050-2214887; e-mail: cicogna@df.unipi.it)}

\newtheorem{theorem}{Theorem}
\newtheorem{lemma}{Lemma}
\newtheorem{proposition}{Proposition}

\newtheorem{definition}{Definition}

\def \ov{\over}
\def \bar{\overline}
\def \beq{\begin{equation} }
\def \eeq{\end{equation} }
\def \lb{\label}

%%%%%%% notaz matem
\def \pd{\partial}
\def\~#1{\widetilde #1}

%%%%%%%% abbrev
\def \sy {symmetry}
\def \sys {symmetries}
\def \so {solution}
\def \eq{equation}
\def \R{{\bf R}}

%%%%%%%% greco
\def\a{\alpha}
\def\be{\beta}
\def\phi{\varphi}
\def\De{\Delta}
\def\la{\lambda}

\def\Na2{\nabla^2}

\def \qq{\qquad}
\def \q{\quad} 
\def\={\, =\, }

\baselineskip .5cm
\date{}
\maketitle

\begin{abstract}
\noindent
We consider the problem of performing the preliminary ``symmetry classification'' 
of  a class of quasi-linear PDE's  containing one or more arbitrary functions:  we 
provide an easy condition involving these functions in order 
that nontrivial Lie point symmetries be admitted, and a ``geometrical'' 
characterization of the relevant system of equations  
determining these symmetries. Two detailed  examples will  elucidate the idea
and the procedure: the first one concerns a  nonlinear Laplace-type equation, the 
second a generalization of an equation (the Grad-Schl\"uter-Shafranov  equation) 
which is used in magnetohydrodynamics.
\end{abstract}


\bigskip\bigskip

\centerline{{\sl Running title}: Symmetry Classification  of  quasi-linear PDE's}

\vfill\eject

\section{Introduction}

The analysis of \sy\ properties of differential \eq s is a well 
established and widely used tool both for studying general properties of 
the \eq s and for finding their \so s (see e.g. \cite{Ov}-\cite{BA} and references 
therein); actually, the determination of Lie-point \sys\ (we will
consider  only \sys\ of this type) is by now an almost completely standard routine, 
thanks  also to some suitably dedicated computer packages 
(see e.g. \cite{CHW}-\cite{Ba}). 

The situation is however considerably different when one deals with the 
problem of performing the ``\sy\ classification'' of an \eq\ which
contains one  or more arbitrary functions, and one wants to discover how
the \sy\  properties depend on the choice of these functions: the
discussion may be far from easy, as far as the \sys\ may drastically
change when the functions  are changed: see e.g. \cite{PY}-\cite{ZL} and references therein.

In this paper, we discuss the case of PDE's of the form given below 
(eq.(\ref{i})) containing one or more arbitrary functions $F_\ell(u)$ of the
unknown variable $u$, and we  shall give an easy condition
involving these functions in order  that the \eq\ admits nontrivial \sys .
This will provide a neat ``geometrical''  characterization of the relevant system of
\eq s   determining  these \sys , and a direct way to perform the complete \sy\ classification of the given \eq  . Two examples will  elucidate the idea and the procedure. 
The first one concerns a   nonlinear Laplace-type \eq ; the 
second example deals with a generalization of an equation (the Grad-Schl\"uter-Shafranov  
equation) which is used in magnetohydrodynamics and plasma physics \cite{Wes}.

\section {Preliminary results}

For the sake of simplicity we consider only second order \eq s for the
unknown function $u=u(x,y)$ of two independent  variables $x,y$  (but the
extension to more general cases is completely  straightforward),
and we will deal with quasi-linear PDE's of the following form
 \beq
 a_{11} \, u_{xx}+a_{12} \, u_{xy}+a_{22} \,
u_{yy}+  b_1 \,  u_x+b_2 \, u_y= 
 \sum_{\ell=1}^L \a_\ell \, F_\ell(u) \lb{i} 
\eeq
or, in a short-hand notation,
\[{\cal E}[u]\=  \a_\ell \, F_\ell(u)\]
where $a_{ij}=a_{ij}(x,y) ,\, b_i=b_i(x,y)  ,\, \a_\ell=\a_\ell(x,y)$ are given (smooth) functions, and
$F_\ell(u)$ are $L$ arbitrary (smooth) functions of $u$ (in the
examples below we will deal with just one or two functions
$F_\ell(u)$). It is understood that no linear relations exist  between $\a_\ell$ and between $F_\ell$. We will  exclude from our consideration the rather 
trivial case when $F_\ell(u)$ are linear functions of $u$, which usually 
can be more simply and conveniently discussed separately by means of direct calculations. 
In this paper we are looking only for Lie-point \sys\ of (\ref{i}); 
we will denote their Lie generator by
\[
X\=\xi(x,y,u){\pd\over{\pd x}}+\eta(x,y,u){\pd\over{\pd
y}}+\phi(x,y,u){\pd\ov{\pd u}}\, .
\]
Assume e.g.  $a_{11}\not=0$ (and put then $a_{11}=1$). 

Although essentially standard (see  \cite{Ov,Ol,BA}, and also \cite{I2} for some generalization), let us summarize for convenience and in order to fix notations,  these first basic results, which can be easily obtained imposing the usual \sy\ condition
\[
 X^{(2)}\,(\De)\Big|_{\De=0}\ =\ 0   \]
where  $\De={\cal E}[u]-  \a_\ell F_\ell$, 
and $X^{(2)}$ is the second prolongation of $X$.
\begin{lemma}  For any choice of the functions $F_\ell$, the 
coefficients of the 
Lie-point \sy\ operators admitted by a PDE of the form (\ref{i}) satisfy the 
conditions $\xi_u=\eta_u=0 ; \, \phi_{uu}=0$, i.e.
\beq
 \xi=\xi(x,y)\q  ,\q \eta=\eta(x,y)\q ,\q  \phi=A(x,y)+uB(x,y)\eeq
and the unique determining \eq\ which involves the functions $F_\ell(u)$ takes the form
(with $F'_\ell=dF_\ell/du$, sum over $\ell=1,\ldots,L$) 
\begin{gather} \lb{pf}
  p_\ell(x,y)F_\ell(u)+  
 p_{L+\ell}(x,y)F'_\ell(u)+    p_{2L+\ell}(x,y)uF'_\ell (u)+ \\  p_{3L+1}(x,y)\, u+
  p_{3L+2}(x,y)  =  0 \notag
\end{gather}
where the coefficient  functions $p_i(x,y)\ (i=1,\ldots,3L+2)$ are given by 
\begin{gather}
p_\ell=B\a_\ell-(\xi_x+\eta_y)\a_\ell-a_{12}\xi_y\a_\ell-\xi\a_{\ell,x}-
\eta\,\a_{\ell,y}\ ,   \lb{pp} \\
p_{L+\ell}=-\a_\ell A \  , \  p_{2L+1}=-\a_\ell B \notag \  ,\  
p_{3L+1}= {\cal E}[B] \  ,\   
p_{3L+2}={\cal E}[A]\notag
\end{gather}
with $\xi_x=\pd\xi/\pd x, \ \a_{\ell,x}=\pd\a_\ell/\pd x$ etc. $(\ell=1,\ldots,L$).
\end{lemma}
Considering now the determining \eq\ (\ref{pf}), and observing 
that the $p_i$ depend only on $x,y$ and not on $u$, one immediately 
realizes that, if the $3L+2$ functions $f_i$ defined by
\beq\lb{fL}
f\equiv(F_\ell,F'_\ell,uF'_\ell,u,1) \q\q\q (\ell=1,\ldots,L)\eeq
are {\it linearly independent}, then (\ref{pf}) can be satisfied if and
only if
\beq\lb{p0} p_i\ =\ 0 \q\q (i=1,\ldots , 3L+2) \ . \eeq
Recalling now the definition of {\it kernel} of the full (or principal) \sy\ groups \cite{Ov} of eq. (\ref{i}), i.e. the intersection of all \sy\ groups admitted by (\ref{i}) for any arbitrary choice of $F_\ell(u)$, we can then state the following property.
\begin{lemma} Conditions (\ref{p0}) characterize the kernel  of the  \sy\ groups of \eq\ (\ref{i}).
\end{lemma}
Indeed, conditions (\ref{p0}), together with the other determining \eq s (not involving $F_\ell$), determine
the functions $\xi,\eta,A,B$ (i.e. the  \sys\ admitted by   \eq\
(\ref{i})), which are independent of the 
the choice of the functions $F_\ell$. These \sys\ may be considered
``trivial'' in this context: for instance, if all coefficients $a,b,\a$ in
(\ref{i}) are independent of $y$, then such a \sy\ operator is $\pd/\pd y$. Some not so obvious
examples of \sys\ of this type will be presented later.

Therefore, a first conclusion is that, in order to have
``nontrivial'' \sys\ (i.e. really dependent on the choice of the functions $F_\ell$), 
a necessary condition is the existence of some linear dependence
among the functions~(\ref{fL}).  

Another relevant remark which will emerge from our discussion is the important role played also by the coefficient functions $\a_\ell(x,y)$ in the determination of the admitted \sys .

\section {Conditions for the existence  of \sys .}

Consider the linear space generated by the $3L+2$ functions $f_i$ defined
in (\ref{fL}), and, according to our above remarks, now assume that there are 
some linear relations among
these functions. Then the $f_i$ span a space with dimension $k<D=3L+2$; if
this is the case, the $D$ coefficients $p_i$ are forced, according to
(\ref{pf}), to belong to the {\it orthogonal} $(D-k)$-dimensional subspace
(with respect to the standard scalar product in $\R^D$), and the functions
$p_i(x,y)$ turn out to be subjected to $k$ linear conditions. For
instance, if there is just one linear relationship between the $f_i$, say
\beq\lb{lf}
 \sum_{i=1}^{D}\la_i\, f_i\ =\ 0 \eeq
where not all the constants $\la_i$ vanish, then $k=D-1$ and the functions $p_i$ span a
1-dimensional subspace and must satisfy  
$D-1$ \eq s of the form (assuming that, e.g., $\la_D\not=0$)
\beq\lb{pla}
p_1\la_D  = \la_1p_D\   ,\  p_2\la_D  =  \la_2p_D\   ,\  \ldots \, , \ 
p_{D-1}\la_D  =  \la_{D-1}p_D \ . \eeq
We can then state our main conclusion, which  characterizes the crucial 
determining \eq\ which contains the functions $F_\ell$ in the following ``geometrical'' form.
\begin{proposition}  Eq. (\ref{i}) 
admits nontrivial \sys\ only if the $D=3L+2$ functions (\ref{fL}) 
are linearly dependent. If this is the case, the $D$ functions $p_i(x,y)$ given by   (\ref{pp})
appearing in the determining \eq\ (\ref{pf}) span the subspace orthogonal 
(with respect to the standard scalar product in $\R^D$) to the $k-$dimensional 
($k<D$) subspace spanned by the functions (\ref{fL}). The admitted \sys\ are completely determined by imposing this orthogonality condition to the coefficient functions $p_i(x,y)$, together with the other determining \eq s not involving the functions $F_\ell(u)$.
\end{proposition}

Before considering explicit examples, let us remark that the complete \sy\ classification must be accompanied by the determination of the equivalence group \cite{Ov}, i.e. the group of the transformations  which leave invariant the differential structure of the PDE. Standard calculations show easily that, for any fixed choice of the functions $a_{ij},\ b_i,\ \a_\ell$ in~\eq~(\ref{i}), the equivalence group  includes in particular,  expectedly,  the scalings $u\to c\,u,\, F_\ell\to c\,F_\ell$  and the translation $u\to u+c_0$ ($c,\,c_0=$ const.). Other transformations involving also the variables $x,y$ can appear for particular choices of the functions $a_{ij},\, b_i,\, \a_\ell$. The   transformations belonging to the equivalence group will play an important role in performing the complete \sy\ classification of our \eq s.

%\vfill\eject

\section{First example: a generalized Laplace \eq }
To illustrate the main idea and the procedure, and also to clarify some details, we are now going to examine some examples, which can be noteworthy also for their different and interesting peculiarities. We start considering the simplest case, where the r.h.s. of (\ref{i}) contains only one
term $\a(x,y)F(u)$ (then $D=5$).

First of all, let us remark that, independent of the form of the \eq\ (\ref{i}), 
not all linear combinations among the $f_i$ defined in (\ref{fL}) can be  chosen
arbitrarily. For instance, no relation exists between $u$ and $1$, and
also, having excluded the trivial case  of  linear $F(u)$,
between $F,\, u,\, 1$. On the other hand, the necessary linear dependence between the functions $f_i$ implies immediately that the presence of some \sy\ is possible only if $F(u)$ is an exponential or a power of $u$.  It can also happen that more than one linear relation  holds:  e.g., if  $F=u^2/2+u$, then  both
relations
\beq\lb{uF}
F'-u-1\ =\ 0 \q {\rm and}\q 2F-u\,F'-u\ =\ 0 \eeq
hold simultaneously, i.e. $k=3$.  In this case, the orthogonality condition expressed by Proposition 1 becomes
\beq\lb{p4} p_1+2p_4-2p_5=0\  ,\  p_2+p_5=0\  ,\  p_3-p_4+p_5=0 \ .\eeq

As an explicit example, let us  consider the following  generalization of the classical nonlinear Laplace \eq 
\beq\lb{Li}
\De\equiv\Na2 u-\a(x,y)F(u)\ =\ 0   \eeq
where $\a(x,y)$ is a given function.
In this case,  the determining \eq s not containing $F$ imply in particular
\beq\lb{hc} \xi_x=\eta_y\q ,\q \xi_y=-\eta_x\q,\q   B={\rm const} \eeq
whereas the crucial determining \eq\ (\ref{pf}) involving $F$ is
\beq\lb{deteq}  p_1F+p_2F'+p_3uF'+p_4u+p_5\=0\eeq
with the coefficient functions $p_i(x,y)$ given by
\begin{gather}
p_1=B\a-(\xi_x+\eta_y)\a-(\xi\a_x+\eta\a_y)\  , \ p_2=-\a A\ , \lb{px}\\
 p_3=-\a B \ , \ p_4=\Na2 B=0 \ , \ p_5=\Na2 A\ . \notag
\end{gather}

Let us first discuss the kernel group.
The conditions $p_i=0\ (i=1,\ldots,5)$ characterizing the transformations 
in the kernel group (see Lemma 2)
give now $A=B=\phi=0$ and the condition $p_1=0$, which now reads 
$\a(\xi_x+\eta_y)+(\xi\a_x+\eta\a_y)=0$. 
Introducing a harmonic function $\Phi=\Phi(x,y)$ such that
\[ \Phi_x\=\xi \q\q\q \Phi_y\=-\eta\]
this condition can be more conveniently  transformed into an \eq\ for the single unknown $\Phi$:
\beq\lb{Phi}
\a(\Phi_{xx}-\Phi_{yy})+\a_x\Phi_x-\a_y\Phi_y\=0 \ .
\eeq
Solutions of this \eq\ clearly depend on the choice of the function $\a(x,y)$.  For instance, if $\a=\,$const.,   it gives $\xi_x+\eta_y=0$,
which, together with (\ref{hc}), implies that the   \sys\ in the kernel group are, as expected, only translations and rotations of the variables $x,y$. If $\a=\exp(2x)$ then $\Phi=\exp(-x)\big(c_1\cos y+c_2\sin y\big)+c_3 y+c_4$ and the kernel contains, apart from the translation generated by $\pd/\pd y$, the transformations generated by
\[ X_1=\exp(-x)\Big(\cos y{\pd\ov{\pd x}}-\sin y{\pd\ov{\pd y}}\Big) \q , \q
X_2=\exp(-x)\Big(\sin y{\pd\ov{\pd x}}+\cos y{\pd\ov{\pd y}}\Big) \ .\]
With $\a=x^r$,   one has that if $r\not=-2$ then the kernel contains only the translation generator
$\pd/\pd y$, whereas if $r=-2$ it also contains the transformations generated by the two operators
\[ X_1\= 2xy{\pd\ov{\pd x}}-(x^2-y^2){\pd\ov{\pd y}} \qq {\rm and}\qq X_2\= x{\pd\ov{\pd x}}+y{\pd\ov{\pd y}} \ . \]
The first transformation  describes the kernel group even if $\a=x^{-2}\be\big(y/(x^2+y^2)\big)$, 
where $\be$ is an arbitrary function.

It can be remarked, incidentally, that if we reverse  the argument for a moment,  one has that:  given {\it any} harmonic function $\Phi$ (and then any couple of harmonic conjugate functions $\xi,\,\eta$), there are some $\a(x,y)$ which solve  \eq\ (\ref{Phi}), and then, with these functions $\a$, the kernel group contains precisely the \sy\ generated by the operator $X=\xi(\pd/\pd x)+\eta(\pd/\pd y)$.

Let us now finally perform the \sy\ classification of eq. (\ref{Li}).  As  already remarked, its equivalence group may contain, in addition to the transformations listed at the end of Section 3, and depending on the specific choice of the function $\a$, other transformations  possibly  involving also $x,y$. As we shall see, however, these are not relevant for the \sy\ classification of \eq\ (\ref{Li}). 

According to our procedure, it is immediately seen that just {\it one} linear relation between the five $f_i$ can exist. For instance,
in the case $F=u^2/2+u$ mentioned above,   admitting the two linear relations (\ref{uF}), conditions (\ref{p4}) would lead to $p_i=0$, i.e. only the  kernel \sys . 
We then assume the existence of a single linear relation:
\beq\lb{LL}  
\la_1F+\la_2F'+\la_3uF'+\la_4u+\la_5\=0\eeq
with not all $\la_i$ equal to zero. Observing that $p_4=0$, one has from (\ref{pla}) that $\la_4=0$. 
We now distinguish the cases $\la_3\not=0$ and $\la_3=0$.

Let $\la_3\not=0$. It is not restrictive to put $\la_3=1$, and (up to a translation of $u$) $\la_2=0$, 
which implies   $\la_5\,p_2=0$. If $\la_5\not=0$, then 
$p_2=0$ would imply $A=0$ and also $p_5=0=p_3\la_5$; now, if $p_3=0$ it 
remains only  $p_1\not=0$, and from $\la_5p_1=\la_1p_5=0$ one concludes 
that $\la_5=0$. Therefore, we get $ \la_1F+uF'\=0$, i.e.
\[ F(u)\=u^m \q{\rm with} \q m\=-\la_1 \q\q (m\not=0,\, 1) \ .\]
Using now (\ref{pla}), which become $ p_1+mp_3=0\ ,\ p_2=p_4=p_5=0$,
we get
\[ A=0\q   {\rm and} \q \a B(m-1)+(\xi_x+\eta_y)\a+(\xi\a_x+\eta\a_y)=0 \ . \]
The last \eq\ relates the \sy\ coefficients $\xi,\,\eta,\,B$ with the specific form of the function $\a(x,y)$. 
If for instance $\a=x^r$, then $B(m-1)+\xi_x+\eta_y=0$, but $B$ must be $\not=0$, otherwise also $p_3=p_1=0$, i.e. the kernel group. Therefore, $\xi$ must be proportional to $x$ and   \eq\ (\ref{Li}) admits the \sy\ operator
\[ X\=  (m-1)\Big(x{\pd\ov{\pd x}}+y{\pd\ov{\pd y}}\Big)-(r+2)\,u{\pd\ov{\pd u}}  \]
(and obviously the translation of the variable $y$, and also the translation of $x$ and the rotations 
of $x,y$ in the case $r=0$, i.e. if $\a=$ const.).

Let now $\la_3=0$. Then necessarily $\la_2\not=0$, and one can put $\la_2=1$ and also $\la_1=1$ (possibly up to a scaling of $u$). Assume first $\la_5=0$; therefore, from (\ref{LL}),
\[ F(u)\= \exp(-u) \]
and the conditions (\ref{pla}),(\ref{px}) for the functions $p_i$ become now
\[  p_3=-\a B=0 \ , \ p_5=\Na2 A=0\ , \ {\rm and}\ 
p_1=p_2\q {\rm i.e.}\q  \]
\[ (\xi_x+\eta_y)\a+(\xi\a_x+\eta\a_y)=\a A \ .\]
As before, we can consider some examples. If $\a=$ const.,  the last \eq\ shows $A=\xi_x+\eta_y$ and  then the most general \sy\ of the \eq\ $\Na2 u=\exp(-u)$ is 
\[ X\= \xi{\pd\ov{\pd x}}+\eta{\pd\ov{\pd y}}+(\xi_x+\eta_y){\pd\ov{\pd u}} \]
where $\xi,\,\eta$ are arbitrary harmonic conjugate functions: this is the well known case of the classical Liouville \eq\   (in its ``elliptic'' form; for a full discussion of the \sy\ properties and other related features of the Liouville-type \eq s, see \cite{FuS}-\cite{Br}). If instead for instance $\a=y^r$ then the admitted \sy\ operators are
\[ X_1=x{\pd\ov{\pd x}}+y{\pd\ov{\pd y}}+(2+r){\pd\ov{\pd u}} \q  , \q X_2= (x^2-y^2){\pd\ov{\pd x}}+2xy{\pd\ov{\pd y}}+(2r+4)x{\pd\ov{\pd u}}   \]
and the translation $\pd/\pd x$.
Assume now $\la_5\not=0$, then
\[ F(u)\= \exp(-u)-\la_5 \ . \]
Introducing the transformation $u\to u+\~u$, where $\~u=\~u(x,y)$ 
satisfies the \eq\ $\Na2\~u+\la_5\a=0$, one obtains the new \eq
\[ \Na2 u-\~\a\exp(-u)\=0 \]
where $\~\a(x,y):=\a\,\exp(-\~u)$, which has precisely the same form as the \eq\ 
considered before. Without repeating details, it can be interesting to 
provide just one illustrative example. Let
\[ \Na2 u- k\,x^{-2} \big(\exp(-u)+1\big)\= 0 \]
where $k=$ const. It is easy to see that if $k=2$ this \eq\ admits the \sys
\[ X\=\xi{\pd\over{\pd x}}+\eta{\pd\over{\pd 
y}}+\Big(\xi_x+\eta_y-{2\over x}\xi\Big){\pd\over{\pd u}} \]
where $\xi,\, \eta$ are arbitrary harmonic conjugate functions; if instead 
$k\not=2$, the admitted \sys\ are only those in the kernel group.

The above results concerning  eq. (\ref{Li}) can be stated in a complete form as follows.
\begin{proposition}
Given a function $\a=\a(x,y)$,  consider this \eq\ for the harmonic function $\Phi=\Phi(x,y)$
\beq\lb{hcc}\a(\Phi_{xx}-\Phi_{yy})+\a_x\Phi_x-\a_y\Phi_y -\a C \= 0 \ . \eeq
Let $\xi=\Phi_x,\,\eta=-\Phi_y$. Assume first $C=0$: for any \so\ $\Phi$ of (\ref{hcc}),   the kernel group of the generalized Laplace \eq\ (\ref{Li}) is generated by the \sy\ operator
\[ X\= \xi{\pd\ov{\pd x}}+\eta{\pd\ov{\pd y}}\ . \]
If $F(u)=u^m$, for any \so\ $\Phi$ of (\ref{hcc}) with $C= $ const.$\not=0$,  eq. (\ref{Li}) admits the \sy\
operator
\[ X\=(m-1)\Big( \xi{\pd\ov{\pd x}}+\eta{\pd\ov{\pd y}}\Big)- Cu{\pd\ov{\pd u}} \ .\]
If $F(u)=\exp(-u)$, for any \so\ $\Phi$ of (\ref{hcc}) with $C=C(x,y)$ any  harmonic function, eq. (\ref{Li}) admits the \sy\ operator
\[ X\= \xi{\pd\ov{\pd x}}+\eta{\pd\ov{\pd y}}+C(x,y){\pd\ov{\pd u}} \ . \]
In particular,  if $\a=$const.  then $C=\xi_x+\eta_y$ where $\xi,\,\eta$ are arbitrary harmonic conjugate functions,  and  the case of the standard ``elliptic'' Liouville \eq\ is recovered. If finally 
$F(u)=\exp(-u)+c$ ($c=$ const.), the above case is recovered by 
means of the transformation $u\to u+\~u$, where $\~u=\~u(x,y)$ 
satisfies the \eq\ $\Na2\~u-c\,\a=0$.
This completes the \sy\ classification of the PDE (\ref{Li}), apart from the transformations in the equivalence group.
\end{proposition} 

\section {An example with two arbitrary functions.}

We now consider the case of a PDE of the form (\ref{i}) with two arbitrary functions $F_\ell(u)$, i.e. 
$L=2, \ D=8$. To avoid excessive generality, let us restrict our study to a PDE of the following form
\beq \lb{GSS}
u_{xx}+u_{yy}+{a\ov x}u_x\=\a(x,y)F_1(u)+F_2(u)\eeq
here $b_1=a/x,\, a\not=0$ is a constant,  $\a_1=\a(x,y)$ a given function and $\a_2=1$. The choice of this \eq\ is motivated and suggested by the theory of plasma physics: it is indeed a generalization of the 
Grad-Schl\"uter-Shafranov equation (see \cite{Wes}), which is obtained putting in (\ref{GSS}) 
$a=-1,\ \a=x^2$,  and describes  the  magnetohydrodynamic force balance in
a magnetically confined  toroidal plasma. In this context,
$u$ is the so-called magnetic flux variable,  $x$ is a radial  variable
(then $x\ge 0$), while the two arbitrary  functions $F_1(u)$, $F_2(u)$ 
are flux functions related to the plasma pressure and current density profiles.

The determining \eq s not involving $F_\ell(u)$ give in this case   
\[ \xi_x=\eta_y,\  \xi_y=-\eta_x, \q {\rm and} \q B={-a\xi\ov{2x}}+b, \q b={\rm const}  \ .\]
First of all, the kernel group is immediately seen to be trivial (apart obviously from the translation generated by $\pd/\pd y$, in the case where $\a$ depends only on $x$; the possible presence of this \sy\ will be tacitly understood in the following). Indeed, from $p_i=0$ (see  Lemma 2), 
one has $A=B=0$,  then the above \eq\ implies   $\xi=(2b/a)\, x$, and the condition $p_2=0$ 
with $\a_2=1$ gives finally $\xi=0$.

Let us now start assuming that there are exactly {\it two} linear relations involving the functions $F_1$ and $F_2$ separately:
\begin{gather} 
\la_1 F_1+\la_3F'_1+\la_5uF_1'+\la_7u+\la_8\=0   \lb{lf2}\\
\la_2 F_2+\la_4F'_2+\la_6uF'_2+\mu_7u+\mu_8\=0 \notag
\end{gather}
with not all $\la_i,\mu_i$ equal to zero. 

Let $\la_5\la_6\not=0$, and put $\la_5=\la_6=1$. According to Proposition 1,   the \sy\ coefficients $p_i$, given by (\ref{pp}),  satisfy then the six linear conditions
\begin{gather}
 p_1=\la_1p_5 \  , \  p_2=\la_2p_6 \  ,\  p_3=\la_3p_5 \  ,\  p_4=\la_4p_6\  , \lb{lp2} \\
p_7=\la_7p_5+\mu_7p_6 \  ,\  p_8=\la_8p_5+\mu_8p_6  \notag
\end{gather}

With $\la_5\not=0$, we can put $\la_3=0$, up to a translation of $u$. Conditions (\ref{lp2}) and the expression of the coefficients $p_i$ give $p_3=0,\, A=0$, then $p_4=p_8=0$ and therefore also $\la_4=\la_8=\mu_8=0$ (indeed, $p_5p_6\not=0$, otherwise all $p_i=0$); condition  $p_2=\la_2p_6$   implies that $\xi(x,y)$ must satisfy an \eq\ of the form
\[ \xi_x\= k_0{\xi\ov x}+k_1 \q\q\q (k_0,\, k_1={\rm const}) \]
which admits harmonic \so\ only of the form $\xi=c\,x,\ c=$ const. On the other hand, $\la_1=p_1/p_5,\ \la_2=p_2/p_6$   imply that $\a$ is forced to satisfy
\beq \lb{ar} {x\a_x+y\a_y\ov {\a}}\=r\={\rm const} \ .\eeq
This means that if $\a$ does not satisfy this condition,  no \sy\ is allowed; we then assume for $\a$ the form
\[ \a(x,y)=x^r\,\be(y/x)\]
where $\be$ is arbitrary. Notice that, with  $\a$ of this form, a new transformation is included in the equivalence group, namely the scaling $x\to cx,\ y\to cy, F_1\to c^{2-r}F_1,\ F_2\to c^{-2}F_2$. We also deduce 
$B=$ const.~$\not=0\, ,\, p_7=0$, which implies in turn $\la_7=\mu_7=0$. Then we are left with
\[ \la_1F_1+uF'_1\= 0 \q , \q \la_2F_2+uF_2'\= 0 \]
giving (thanks to some scalings -- all these transformations belong indeed to the equivalence group)
$ F_1=u^{-\la_1}, \ F_2=u^{-\la_2} $ where
\[- \la_1\= 1-{c\ov B}(2+r)\= 1+{2+r\ov q}  \q ,\q -\la_2\= 1+{2\ov q}  \q {\rm with}\q B\=-c\,q \]
with admitted \sy\ generated by
\[ X\= x{\pd\ov{\pd x}}+y{\pd \ov{\pd y}}-q \, u{\pd\ov{\pd u}} \ . \]

Let now $\la_5=\la_6=0$. Then necessarily $\la_3\la_4\not=0$. According to Proposition 1, the orthogonality condition now reads (with $\la_3=\la_4=1$)
\[ p_1\=\la_1p_3 \  ,\  p_2\=\la_2p_4 \  ,\  p_5=p_6\=0 \  , \ 
p_7\=\la_7p_3+\mu_7p_4 \  , \  p_8\=\la_8p_3+\mu_8p_4 \ .\]
In this case, one has immediately $B=0$, then $p_7=0$ and $\la_7=\mu_7=0$,  and again $\xi=c\,x$. From $p_2=p_4$ one has $2\xi_x=A=$ const., which 
gives $p_8=\la_8=\mu_8=0$. Up to a scaling of $u$, one can choose $\la_2=1$, the \eq s for 
$F_1,\, F_2$ are then 
\[ \la_1F_1+F_1'\=0 \q ,\q F_2+F_2'\=0 \]
giving  $F_1=\exp \big(-\la_1u\big)\, ,\, F_2\=\exp\big(-u\big)$, and finally from $\la_1=p_1/p_3$ one deduces the same condition (\ref{ar}) as before for the function $\a(x,y)$, and $\la_1=1+(r/2)$.

The conclusion will be stated in complete form in the following Proposition 3. Indeed,
with the same, and even simpler, arguments used in the two above cases, it is an easy task to see that no other possibilities are left to  the PDE (\ref{GSS}) of admitting \sys .
\begin{proposition}
The kernel group of \eq\ (\ref{GSS}) is trivial (apart from the translation $y\to y+c$ if $\a$ depends 
only on $x$). Except for this, a necessary condition in order that  \eq\ (\ref{GSS}) may admit \sys\ is that the function $\a$ has the form $\a(x,y)\= x^r\be(y/x)$, where $\be$ is an arbitrary function. 
With  $\a$ of this form, \eq\ (\ref{GSS})  admits  a \sy\ 
only with the following choices for  the functions $F_1(u),F_2(u)$ (up to transition
to equivalent functions via equivalence group): 
\noindent
\[\hspace*{-2.9cm}\mbox{$a)$} \quad\q\q\q\q F_1(u)\ =\ u^{1+(r+2)/q } \quad , \quad F_2(u)\ =\ u^{1+(2/q)}\]
for all $q\not=0$, with  admitted \sy\ operator
\[
X=x\frac{\partial}{\partial x}+y\frac{\partial}{\partial y}- \,q\,u\,
\frac{\partial}{\partial u} \]
and
\[ \hspace*{-2.0cm} \mbox{$b)$} \quad\q\q\q\q
F_1(u)=\exp \Big(\! -(1+{r\ov 2})\, u\Big) \quad ,\quad F_2(u)=\exp (-u) \]
with \sy\  operator
\[ X=x\frac{\partial}{\partial x}+y\frac{\partial}{\partial
y}+2\frac{\partial}{\partial u} \ . \]
\end{proposition}
As stated at the beginning of Section 2, we have excluded from our analysis the case where the functions  $F_1$ and $F_2$ are linear functions of $u$. This case indeed, in general of minor interest,  can be more conveniently considered by means of  a separate and direct calculation. Specifically, if the r.h.s. of \eq\ (\ref{GSS}) has the form $\~\a(x,y)u+\~\be(x,y)$, the admitted \sy\ is, not surprisingly, generated by
\[ X\=\big(c\,x+\Psi(x,y)\big){\pd\ov{\pd u}}\]
where $c$ is a constant and $\Psi(x,y)$ is any \so\ of the PDE
\[{\cal E}[\Psi]\=\~\a\Psi-c\,\~\be \ .\]

In the particular case of the Grad-Schl\"uter-Shafranov \eq\ of plasma physics \cite{Wes}, the above results were already presented  in Ref.~\cite{Sig}, but without any details in the calculations and without any reference to the procedure, which is instead one of the main purposes of the present paper. In the same reference~\cite{Sig} one can also find some physical comment on the  \sy\ properties of the above \eq .

\section*{Acknowledgments} 

The author is gratefully indebted to the reviewers for 
their care in reading the manuscript, and for their valuable 
suggestions and helpful comments.

\label{lastpage}

\end{document}